\documentclass[12pt]{article}
\usepackage{amsmath} \usepackage{amssymb}
 \newcommand{\fns}{\footnotesize}
 \newcommand{\beq}{\begin{equation}}\newcommand{\eeq}{\end{equation}}
\newcommand{\ben}{\begin{enumerate}}\newcommand{\een}{\end{enumerate}}
 \newcommand{\Ups}{\Upsilon} \def\Lms{\Lambda_{\msbar}}
\newcommand{\eps}{\epsilon}
\newcommand{\veps}{\varepsilon}
\newcommand{\ac}{\mathcal{A}}
\def\al{\relax\ifmmode\alpha\else{$\alpha${ }}\fi}
\def\alps{\relax\ifmmode\alpha_s\else{$\alpha_s${ }}\fi}
\def\as{\relax\ifmmode\alpha_s\else{$\alpha_s${ }}\fi}
\def\albars{\relax\ifmmode{\bar{\alpha}_s}\else{$\bar{\alpha}_s${ }}\fi}
\def\agoth{\relax\ifmmode{\mathfrak A}\else{$\,{\mathfrak A}${ }}\fi}
 \def\agothk{\relax\ifmmode{\mathfrak A}_k\else{${\mathfrak A}_k${ }}\fi}
\def\agothks{\relax\ifmmode{\mathfrak A}_k(s)\else{${\mathfrak A}_k(s)${}}\fi}
\def\acal{\relax\ifmmode{\cal A}\else{${\cal A}${ }}\fi}
  \def\acalk{\relax\ifmmode{\cal A}_k\else{${\cal A}_k${ }}\fi}
 \def\msbar{\relax\ifmmode\overline{\rm MS}\else{$\overline{\rm MS}${ }}\fi}
\def\eV{\relax\ifmmode{\rm e\kern-0.12em V}\else{\rm e\kern-0.12em V{ }}\fi}
\def\MeV{\relax\ifmmode{\rm M\eV}\else{\rm M\eV{ }}\fi}
\def\GeV{\relax\ifmmode{\rm G}\eV\else{\rm G\eV{ }}\fi}            
\begin{document}
\begin{center}
 \large{\bf Analytic Perturbation Theory Model \\ for \ QCD \
   and \ Upsilon \ Decay}\\
 \bigskip
 D.~V.~Shirkov \\
 \medskip{\small \it
  Joint Institute for Nuclear Research, Dubna, 141980,
 Russia }
\end{center}

 \begin{abstract}
 An elegant and more precise (Denominator) formula for
 the 3-loop perturbative QCD coupling is discussed. It
 improves the common expression (e.g., canonized by
 PDG) in few GeV region. 
   On its base, we propose simple analytic Model for
 ghost-free QCD running couplings and their effective
 powers within the Analytic Perturbation Theory, in
 both the space-like (Euclidean) and time-like
 (Minkowskian) regions, very accurate in the range
 above 1 GeV.\par
    Effectiveness of the new Model is illustrated by the
 example of $\Ups(1\mathrm{S})$ decay where the standard
 analysis gives $\alps(M_{\Ups})=0.170\pm 0.004$ value
 that is inconsistent with the bulk of data for $\alpha_s$.
 Instead, we obtain $\as^{Mod}(M_{\Ups})=0.185\pm 0.005$
 that corresponds to $\as^{Mod}(M_Z)=0.120\pm 0.002\,$
 that is close to the world average.\par
\end{abstract}%

 \section{Introduction}
   This text contains essence of three topics
 related to the QCD coupling at low energy:  
 \ben
\item Presenting of QCD perturbative coupling $\as(\mu)$
   by {\sf Denominator representation}
   instead of ``canonical" PDG-like bulky expression.

\item Particular ``Analytic-Perturbation-Theory"(APT)
  ghost-free model for $\as(\mu)$ in the low-energy region.
\item APT Model \ analysis of Upsilonium decay.
  \een
  It is based mainly upon recent paper \cite{Zay05}
 which contain more detailed exposition.

 \subsection{Notation}
  We use the Bethke\cite{Bethke00} notation
(different from the PDG one)
 \beq\label{rg3loop}
 \frac{d\as}{d\,L}=  \beta(\alpha_s)= -\beta_0\,\as^2
 -\beta_1\,\as^3-\beta_2\,\alpha^4 +\,\dots\,,\quad L=
 \ln (x/\Lambda^2)\,,\eeq
  for beta-function coefficients
 $ \beta(\alpha)=-\beta_0\,\alpha^2\left(1+b_1\,\alpha +
 b_2\,\alpha^2+ \dots\right)\,;\, b_k=\beta_k/\beta_0\,,$
{\small $$
\beta_0(n_f)=\frac{33-2\,f}{12\pi}\,;\quad
  b_1(n_f)=\frac{153-19f}{2\pi(33-2f)}\,; \quad
  b_2^{\msbar}(n_f)=\frac{2857-(5033/9)n_f +(325/27)n_f^2}
  {32\pi^2(11-(2/3)n_f)}\, \,$$}
  Here, numerically, all $\beta_k\,,\,b_k\,$ and
 $B=\beta_1^2/\beta_0\,$ are of an order of unity
 {\small $$\beta_0(4\mp1)=0.6631\pm 0.0530\,;\quad
 \beta_1(4\mp 1) =0.3251\pm 0.0802\,; \quad
 \beta_2(4\mp 1)=0.205^{+0120}_{-0.112}\,; $$
 $$b_1(4\pm 1)=0.4902^{-0.0889}_{+0.0757}\,;\,\quad
  b_2(4\mp 1)=0.309^{-0.159}_{+0.144}\,;\quad B(4\mp 1)
  = 0.7392^{+0.0509}_{-0.0814}\,. $$ }\vspace{-5mm}

 \section{Perturbative QCD coupling}
 \subsection{Denominator representation for 3-loop
 $\as(\mu)\,$}
 Instead of canonical {\small( Bethke\cite{Bethke00},
 PDG\cite{pdg04})} $1/L\,$ expanded expression
\[ \albars^{(3)}(x)=\frac{1}{\beta_0L}-
 { \frac{b_1}{\beta_0^2}\frac{\ln L} {L^2}}+
 {\tfrac{1}{\beta_0^3L^3}\left[b_1^2(\ln^2L
 -\ln L-1)+b_2\right]} - \]
 \[.\quad\quad - \tfrac{1}{\beta_0^4L^4}\left[b_1^3
 \left(\ln^3L -\tfrac{5}{2}\ln^2L-2\ln L+\tfrac{1}{2}
 \right) +3b_1b_2\ln L-\tfrac{b_3}{2}\right]+\dots
 \qquad(9.5_{PDG})\]
 we argue to use the {\sf``Denominator representation"}
 for $\as(\mu)\,$
 \beq\label{9.5den}
 \albars^{(3, iter)}(L)=\frac{1}{\beta_0\,L+ b_1\,
 \ln\left(L+ \tfrac{b_1}{\beta_0}\,\,\ln L\right)
 +\tfrac{(b_1^2-{ b_2})}{\beta_0\,L}+ \tfrac{ b_3}
 {2(\beta_0 L)^2}}\,;\quad L=\ln\frac{\mu^2}{\Lambda^2}\eeq
  Canonical form can be obtained from it by use of {\it
  casual} expansion parameter ${\bf\eps_2}\,$
\[ \albars^{(3, iter)}(L)=\frac{1}{\beta_0\,L} \left(1+
 \frac{B\,\ln\left[L+ B\,\ln L\right]}{L} +\veps_2^2-
 \veps_3+ \tfrac{\veps_4}{2}\right)^{-1}\simeq \] 
 \[ \frac{1}{\beta_0\,L}\left[1+({\bf\eps_2}+
 {\bf\eps_2}\veps_2+\veps_2^2)-\left(\veps_3+
 \frac{{\bf\eps_2}^2\veps_2}{2}\right)+ \dots\right]^{-1}
 \,;\quad {\bf\eps_2}=\frac{B\,\ln L}{L}\,\ll\,1\,\]
 as well as the higher loops ones:
 $\veps_{k+1}=\frac{b_k}{(\beta_0 L)^k}\,(k\geq 1)\,.$
 For instance, at $n_f=3,4\,$ 
{\fns \[{\bf\eps_2}(3\div4)={[0.79\div 0.74]\,
 \frac{\ln L}{L}}\,;\quad \veps_2(3\div 4)=
 \frac{[0.79\div 0.74]}{L};\quad  \veps_3(3\div 4)=
    \left(\frac{0.90\div 0.84}{L}\right)^2 ;\quad
 \,\veps_4=\left(\frac{1.2}{L}\right)^3\,.\]}
 Then, ${\bf\eps_2}^2\sim\veps_3\,$ at
 $\ln L^*\sim 1.13\,: L^*\sim 3.3\,,$ that is at
 $ Q^*\sim 5.0\,\Lambda\,,\quad Q\sim 1.6 \div 2\,\GeV\,.$

 There, $\,{\bf\eps_2}\sim 0.25\,,\,\veps_3\sim 0.07\,.$
 Relative $(9.5_{PDG})$ error $\sim {\bf\eps_2}^3\sim 0.02\,,$
 while error of (2) $\sim \veps_4/2\simeq 0.03\,$
  Ghost singularity at $L=0\,$ in (\ref{9.5den}) is
 much weaker than, in $(9.5_{PDG})$.\par  .

\subsection{The beta function arbitrariness}
  In turn, Denominator representation (\ref{9.5den}) is an
 iterative approximate solution of RG differential eq.(1)
 quadrature which, in the two-loop (NLO) case, is
 \beq\label{rg-2}
 \beta_0 L=-\int^{\as}\frac{d a}{a^2(1+b_1\,a)}=\frac{1}
 {\as}+ b_1\,\ln\frac{\as\,\beta_0 } {1+b_1\,\as} \eeq
 -- a transcendental relation that can be resolved\cite{badri00}
 in terms of special Lambert function. Usually, instead,
 one solves this relation iteratively with the result
 \beq\label{iter2}
 \as^{2,iter}(L)\simeq
 \frac{1}{\beta_0\,L+b_1\ln(B+1/\as^{(1)}\beta_0)}=
 \frac{1}{\beta_0\,L+B\beta_0\ln(L+B)}\eeq
 with its famous log-of-log dependence.\par

 Meanwhile, if one transforms 2-loop beta function \'a la
 Pad\'e $\beta_2\,\to \beta_2^P(\alpha) =
 -\frac{\beta_0\,\alpha^2}{1-b_1\,\alpha}\,,$ the result
 of integration and iteration will be different
 \beq\label{rg-2P}
 \beta_0 L=- \int_{}^{\as}\frac{d a}{a^2}(1-b_1\,a)=
 \frac{1}{\as}+b_1\,\ln(\beta_0\,\as)\,,\quad
 \as^{2P,iter}(L)=\frac{1}{\beta_0\,L+b_1\ln L}\,.\eeq
 Thus, the {\it inner self-consistency condition} of the
 2-loop (NLO) approximation looks like
  \beq\frac{\vartriangle\alpha^{(2)}}{\alpha^{(2)}}=
 \frac{\alpha^{(2\,P)}-\alpha^{(2)}}{\alpha^{(2)}}\simeq
 \veps^2_2 \,\ll\, 1\,.\eeq

  Due to this, the intrinsic accuracy estimate for the
 two-loop QCD coupling $\alpha^{(2)}\,$ at 1.5 -- 2 \GeV
 region is about 7 \%, that is (in the \msbar scheme) equal
 to the value of three-loop contribution $\sim\veps_3\,.$
  At the same time, at the b-quark mass $\sim 4.5\,GeV\,$
 one has $\veps^2_2\simeq 0.015\,,\,\,{\bf\eps_2}\sim 0.21\,
 ,\,\,{\bf\eps_2}^3\sim 0.01 \,$\par
 This means that in the $n_f=4,5\,$ domain one can use
 only the first line of the 3-loop eq.$(9.5_{PDG})$ and
 second expression (5) instead of (2) with the same
 accuracy. \par

  Numerically, at 4-flavor region for
 $\Lambda_4\sim 300\,\MeV$ and $ q=1.5;3\,\GeV\,$ one
 has $\veps_2(q=1.5\,\GeV) \sim 0.053\,,\quad
 {\bf\eps_2}^4(1.5)\sim 0.020\,,\veps_2(3)\sim 0.025\,,
 \quad  {\bf\eps_2}^4(3 )\sim 0.020\,$ and at 5-flavors
 for $\Lambda_5\sim 225\,\MeV\,;\quad q\sim 4.5\,\GeV\,$
 -- $ \veps_2(4.5)\sim 0.012\,, \eps^2(4.5)\sim 0.030\,. $

  Pragmatically, this means that, within the 2 per cent
 limit of accuracy, one can equally use simple two-term
 denominator eq.(5) up to $3\,\GeV\,$ scale. However, in
 the $1/L$--expanded form one needs to keep 3 terms (of
 expansion in powers of ${\bf\eps_2}$) in $n_f=5\,$ region
 and 5 terms in the $n_f=4\,$ one. \par

 As it can be shown, the inner consistency of the
 3-loop iterative approximations is controlled by
 the same parameters $\veps_2\,$ and ${\bf\eps_2}\,.$
  At the same time, the measure for importance of direct
 3-loop (NNLO) term $\varepsilon_3=b_2/(\beta_0\,L)^2\,$
 should be compared with $\varepsilon_2\,.$ It turns out
 that they are close (within 40 per cent) to each other.

 \section{ Analytic Perturbation Theory}
\subsection{\large Outline of Analytic Perturbation Theory}  
 Remind first, that the cornerstones of APT are {\it the
 $Q^2$ analyticity of coupling functions} and {\it
 compatibility with linear integral transformations}. For
 the fresh reviews of APT see~\cite{Sh01epjc,Sh05bari}.\par
 Here follows compendium of main definitions.
 The most elegant APT formulation is based on
 the set of spectral functions
 $\{\rho_i(\sigma)\}\,$ defined as
  \beq  \rho_k(z)=\mathrm{Im}([\as(-z)]^k).
  \label{density}\eeq
 The first of them, $\rho_1=\rho(\sigma)\,$ is just
 the K\"allen--Lehmann spectral density for the
 Euclidean APT coupling. Then, higher Euclidean
 (``analyticized $k$th power of coupling in the
 Euclidean domain'') and Minkowskian (``effective
 $k$th power of coupling in the Minkowskian domain")
 APT functions will be
 \beq\label{min} \ac_k(Q^2)=\mathbb{A}[\as^k]=\frac{1}{\pi}
 \int^{+\infty}_0 \frac{\rho_k(\sigma)\,d\sigma}
 {\sigma+Q^2}\,;\quad\agoth_k(s)=\mathbb{R}[\as]
 =\frac{1}{\pi}\int_s^{+\infty}
 \frac{d\sigma}{\sigma}\rho_k(\sigma)\,\,.\eeq
 They are related by integral transformation and satisfy
 differential relations
  \beq\label{r-r}
 \acal_k(Q^2)=\mathbb{D}[\agoth_k]=Q^2\int^{+\infty}_0
\frac{\agoth_k(s)\, d s}{(s+Q^2)^2}\,;\quad
\frac{1}{k}\frac{d\agoth_k(s)}{d\,\ln s}=-
 \sum_{n\geq 1} \beta_{n-1}\agoth_{k+n}(s)\,,\eeq
 which can be used for iterative definitions.


 \subsection{\large Properties of the APT functions}
 For the \underline{\sf one-loop case}, the APT
 formulae are simple and elegant. Starting with the
 perturbative RG-improved QCD coupling $\,\as^{(1)}(Q^2)=
 1/(\beta_0\,l)\,,$ with the help of~(\ref{density}),
 (\ref{min}) one arrives at the ghost-free effective
 Euclidean and  Minkowskian\footnote{Here, we change
 arguments of the APT functions:
 $Q^2\to l=\ln(Q^2/\Lambda^2)\,$ and
 $s\to L=\ln(s/\Lambda^2)\,.$}
 \beq\label{AE11}
 \ac^{(1)}_1(l)=\frac{1}{\beta_0}\left(\frac{1}{l}
 -\frac{1}{e^l-1}\right)\,;\quad
 \agoth^{(1)}_1(L)=\frac{1}{\beta_0\pi}
 \arccos\left(\frac{L} {\sqrt{L^2+\pi^2}}\right)\,\eeq
 APT couplings. Higher functions $\ac_i\,,\agoth_i\,$
 can be defined via recursive relations (\ref{r-r}) with
 only one term in the r.h.s. For instance, Minkowskian
 functions are
 {\small
 \beq\label{2-3}  
 \agoth_2^{(1)}(L)= \frac{1}{\beta_{0}^2}\,
 \frac{1}{L^2+\pi^2}\,, \quad \agoth_3(L)
 =\frac{1}{\beta_{0}^3} \frac{L}{(L^2+\pi^2)^2}\,,\quad
 \agoth_4(L)= \frac{1}{\beta_{0}^4}\frac{L^2-
 \pi^2/3} {(L^2+\pi^2)^3}\,\,.  \eeq }

  \underline{\sf Higher-loop case.} The two-loop
 expressions are more complicated. Here, exact QCD coupling
 \as can be expressed explicitly in terms of a special
 Lambert function $W$ defined as a solution of the
 transcendental equation $W(z)\,e^{W}=z\,.$ This expression
 yields rather complicated formulae for $\alpha_E=\ac_1$
 and $\alpha_Q= \agoth_1$ in terms of the main branch
 $W_{-1}\,$.\par

 At the three-loop case, one meets further complications.
 Here, only for Pad\'e approximated beta-function, exact
 solution can be expressed\cite{badri00} in terms of the
 Lambert function. Such expression are
 not comfortable enough for practical use.\par
  The devised scheme with due account for
 matching\cite{Sh01tmp}, known as ``global APT", has
 been studied numerically by Magradze and Kourashev at
 the two- and three-loop level. They calculated numerical
 tables for the first three functions $\agoth_k{1,2,3}\,$
 and $\acal_{1,2,3}\,$ at three values of
 $\Lambda^{n_f=3}= 350,400,450\,\MeV\,$ in the interval
 $1\,\GeV <\sqrt{s},Q < 100\,\GeV\,$
 ~\cite{badri00,KourMagr01}, and $\agoth_{1,2}\,,\,
 \acal_{1,2}\,$ in the interval $0.1\,\GeV<\sqrt{s},
 Q\lesssim 3\,\GeV\,$~\cite{magr03}.

  The APT functions obey important properties valid
 in the higher-loop case:
 \begin{itemize} 
 \item In the Euclidean and Minkowskian domains,
  QCD couplings $\alpha_E(Q^2)=\acal_1,(Q^2)\,,$
 $\alpha_M(s)=\agoth_1\,$ and their ``effective powers"
   $\acal_k(Q^2)\,$ $\agoth_k(s)\,,$
 are different functions related by integral
 operations $\mathbb{A[\,\,]}$ and $\mathbb{R[\,\,]}\,$
 explicitly defined in eqs. (\ref{min}). 
 Higher functions like (8),(9), are not equal to
 powers of the first ones (\ref{AE11}).

 \item The APT functions $\acal_k(Q^2)\,$ $\agoth_k(s)\,$
 differ of common expansion functions $(\as)^k\,$ in the
 low energy region, where they are regular with finite
 $\alpha_E(0)=\alpha_M(0)=1/\beta_0\,$ or zero limits.
  Unphysical singularities are absent with no additional
 parameters introduced. This behavior provides high
 stability with respect to change of renormalisation
 scheme \cite{ShSolPL97}. In the UV limit, all APT
 functions tends to their usual counterparts $(\as)^k\,.$

  \item Expansion of an observable in coupling powers
 $(\as(Q^2))^n\,$ for the Euclidean or in $(\as(s))^n\,$
 for the Minkowskian case is substituted by nonpower
 expansion in sets $\left\{\acalk\right\}\,,$ or
 $\left\{\agothk\right\}\,$ respectively. The latter
 expansions exhibit a faster convergence.
  \end{itemize}

 \underline{\sf The APT re-examination} of various
 processes has been performed in number of papers~
 \cite{KrasPiv82} -- \cite{BMS05}. In particular,
 paper~\cite{KrasPiv82} contained the first attempt in
 revising the $\Ups$ decay. Simple estimates for the
 influence of $\pi^2-terms$ upon some observables were
 performed in~\cite{Sh01epjc}. The $\tau$ decay was
 re-examined in \cite{Olga03}. Pion form factor was
 studied\cite{BP-KSS04,BMS05} within the APT
 techniques. One more extraction of APT coupling at
 $Q\sim 100-400\,\MeV\,$ was made\cite{Milan02-4}
 by the Milano group from mass spectrum analysis
 of ground and first excited quarkonium states.

\section{Simple Model for 3-Loop APT Functions}
  \subsection{``One-Loop-Like'' Model}    
 Our aim is to construct simple and accurate enough (for
 practical use) analytic approximations for two sets of
 functions $\agoth_k$ and $\ac_k\,,k=1,2,3\,$. To reduce
 number of fitting parameters, one should better provide
 the applicability of the recurrent relations. To this goal,
 we use one-loop APT expressions, eqs.(\ref{AE11}) --
 (\ref{2-3}), with modified logarithmic arguments
 \beq\label{model} 
   \ac^{mod}_k(l)=\ac^{(1)}_k(l_*)\,;\quad
 \agoth^{mod}_k(L)= \agoth^{(1)}_k(L_*)\,,\eeq
 $L_*$ and $l_*$ being some ``two-loop RG times".
  Model functions~(\ref{model}) are related by
 the ``one-loop-type" recursive relations
 \[ \ac^{mod}_{n+1}=-\frac{1}{n\,\beta_0}
 \frac{d\ac^{mod}_n}{dl_*}=-\frac{1}{n\,\beta_0}
 \frac{d\ac^{mod}_n}{dl}\cdot \frac{d l}{d l_*}
 \,,\qquad \agoth^{mod}_{n+1}= -\frac{1}{n\,
 \beta_0}\frac{d\agoth^{mod}_n}{dL_*} \,.\]

 A simple expression for $l_*\,$ was taken from
 \cite{SS-99}, where a plain approximation for the two-loop
 effective log $l_2= l+b\ln\sqrt{l^2+4\pi^2}\,,$ with $b$
 defined in Section 1 was used. This approximation
  combined reasonable accuracy in the low-energy
 range with the absence of singularities for $\alpha_E\,.$
 We extend this approach to higher functions in both the
 Euclidean and Minkowskian domains and change square root
 in ``effective logs" $L_2(a)\,$
 and $l_2(a)\,:\,\sqrt{l^2+4\pi^2}\to\sqrt{l^2+a\pi^2}\,$
 with $a\,,$ an adjustable parameter.
 It comes out from thorough numerical analysis that optimal
 value of the new parameter is $a\approx 2\,,$ while {\it
 effective boundaries between the flavor regions have to be
 chosen on quark masses} $m_c=1.3\,\GeV$ and $m_b=4.3\,\GeV$
 just as in the $\overline{MS}\,$ scheme. \par
  That is, our {\sf Model consists of a set of
 equations (13) with (10) -- (12) and }
 \begin{equation} \label{model2}
   L_*=L_2(a=2)=L+B\ln\sqrt{L^2+2\pi^2}\,,\quad
   l_*=l_2(2)=l+B\ln\sqrt{l^2+2\pi^2}\,.\eeq
  Here, $L\,$ and $l\,$ contain common $\Lms\,$ values, like
 in (\ref{AE11}), for each of the flavor region.
  Advantage of Model (\ref{AE11})--(\ref{2-3}),(\ref{model}),
 (\ref{model2}) is that it involves only one new parameter,
 $a=2$ with $\Lms\,$ and $n_f$ taking their usual values.
\vspace{-3mm}

 \subsection{Accuracy of the Model vs data errors}
   In paper \cite{Zay05}  errors of Model
 expressions (\ref{2-3}), (\ref{model}),(\ref{model2}) in
 each $n_f\,$ range were estimated by numerical comparison
 with the Magradze tables in the interval of 3-loop
 $\Lms^{(n_f=3)} \sim 350 - 400\,\MeV\,.$

  As it follows from that analysis, errors of Model for the
 first three APT functions are small, being of an order of
 1-2 per cent for the first functions, of 3-5\% for the
 second and of 6-10\% for the third ones in the region
 above $1.5\,\GeV\,,$ i.e., in the $n_f=4,5\,$ ranges.
 However, its accuracy in the $n_f=3\,$ region (above
 1\GeV) is at the level of 5-10 per cent.\par

  Meanwhile, relative contributions of typical LO, NLO and
 NNLO terms in APT nonpower expansion for observables
 are usually something like 60-80\%, 30-10\%, and 10-1\%,
 respectively (see Table 2 in Ref.\cite{Sh01epjc}).
  Due to this, the Model accuracy for many cases is defined
 by that of the first model functions $\acal_1^{mod}\,,\,
 \agoth_1^{mod}\,$, provided that QCD contribution to an
 observable starts from one-loop contribution $\sim \as\,.$
   At the same time, for quarkonium decays the leading
 contribution $\sim \as^3\,.$ There, the Model error is
 defined by accuracy of the third
 Minkowskian function $\agoth_3^{mod}\,$.

  In the Table 2 of paper \cite{Zay05} we compared Model
  errors with some data errors in the low energy region.

  With due regard for data errors, we can now set some
 total margin of accuracy that our Model would satisfy.
 This margin may be chosen, e.g., as 1/3 of the data
 error bar, which is no less than 10\%. Then, the accuracy
 limit, imposed on the first APT functions could be about
 3\%, for the second function ca 10\%, and for the third
 ones, about 20\%. Due to this, it is not reasonable to
 use the Model below 0.5 \GeV, whereas above this limit
 it is fully advisable. Now, we proceed to its practical
 application to $\Ups$ decay.\par

 \section{$\Upsilon$(1S) Decay Revised }
  \subsection{$\Ups$  Widths}
 Our first motive for this choice is that the parameter
 $\eps_M(L)=\pi^2/L^2\,$ responsible for deviation of
 APT Minkowskian functions from powers of canonical
 \as, is not very small in the region (5-10 GeV)
 related to this decay: $\eps_M(L)\simeq 0.16 - 0.27
 \,,\quad \eps_M(L_2)\simeq 0.11 -0.18\,.$ \par

  Another processes involving $\Ups\,,$ are $\Ups$
 radiative decays and $\Ups$ production. But they have
 low data precision (about $10\%-40\%$).
  Besides, the non-radiative decay of $1S$-state provides
 the best data precision ($1.5\%$ for the ratio of
 hadronic and leptonic widths~\cite{pdg04}).\par

 The second arguument is the disagreement with the world
 average.

  For an extensive review on $q\bar{q}\,$ decay widths
 see \cite{Bodwin95}. NLO ratio of hadronic and
 leptonic decay widths of $S$ state of the $\Ups$ was
 given in~\cite{Mackenz81}.
  However, their expression is not renorm-invariant.
 In our analysis, to return it to the RG-invariant form,
 we put scale parameter $\mu=M_{\Ups}\,.$ Then
\beq\label{R-Yps}
 R(s_{\Ups})=\frac{\Gamma\left(\Ups\rightarrow hadrons\right)}
 {\Gamma\left(\Upsilon\rightarrow e^+ e^-\right)}=
 \frac{10(\pi^2-9)\as^3(s_\Ups)}{9\,\pi\alpha^2(M_\Ups)}
 \left(1+\frac{\as(s_\Ups)}{\pi}\, 7.2\right)\,,\quad
 s_{\Ups}=M_\Ups^2\,.\eeq
  Then, the issue of scale should be readdressed to
 choice of $s_\Ups\,.$

\subsection{Reevaluation of $\Lambda_{QCD}$
          from $\Ups$ Decay}
 First attempt to evaluate $\Lambda_{QCD}$ extracted from
 $\Ups$ decay by proper taking into account analytic continuation
 effects was made in \cite{KrasPiv82}.  Analogous analysis is
 performed here, employing better accuracy Model expression for
 $\agoth_i(s)$ and using more fresh data from CLEO III
 detector \cite{Adams:2004xa} as well as APT expansion instead of
  formula~(\ref{R-Yps}), that, within our Model,  is reduced to
\begin{equation}\label{UpsMod}
  R_\Ups(s)=5360\left[\agoth_3(L)+
  \frac{\agoth_4(L)}{\pi} \,7.2\right]\,\quad \to \quad
  5360\left(\agoth_3^{(1)}(L_2)+
  2.30\,\agoth_4^{(1)}(L_2)\right)\,. \eeq

 We extracted by this formula $\Lms^{(5)}$ and \as values,
 from fresh data $R_\Ups=37.3\pm 0.75$\cite{Adams:2004xa}.
 In Table , results, obtained within APT are
 compared to  results of standard PT.
\begin{table}[h]  
 \caption{\bf Results of various $\as$ extraction
   from Upsilon decays}
\medskip
\begin{center}
\begin{tabular}{|l|c|c|c|} \hline
\multicolumn{4}{|c|}{Part I. Non-APT treatment}\cr \hline
\multicolumn{1}{|c|}{Source}&$\alpha(M_\Ups)$ &$\alpha(M_Z)$&
$\Lms^{n_f=5}$\cr \hline PDG, $\Ups,1S$
&0.170(4)&0.112(2)&146$^{+18}_{-17}$  \cr \hline PDG, global
Fit&0.182(5)& 0.1185(20)&217$^{+25}_{-23}$\cr \hline\hline
\multicolumn{4}{|c|}{Part II. APT treatment}\cr\hline
 Exact APT,\,1S&0.1805$(12)_{exp}$&0.1179$(5)_{exp}$&210(5)\cr\hline
\bf [Mod],\,$1S$ & \bf 0.185$(5)_{\rm M}$&
 0.120$(2)_{\rm M}$& 235$(25)$\cr\hline
\small\cite{Sh01epjc} Crude APT&0.183 & 0.119 &222\cr\hline
\end{tabular}\end{center}
\end{table}
 In line 1 of Part I, ``PDG, $1S$", standard PT results on
 $\Ups(1S)$ decay are given. We present them not exactly
 as they were published in~\cite{pdg04} but recalculated
 along with modern experimental data. Line 2, marked
 ``PDG, Fit" gives the published world average value
 described by the curve on Fig. 9.2 in \cite{pdg04},
 within the error bars of {\it all} the processes. Column
 ``$\alpha(M_\Ups)$" means ``$\as$,  calculated at the
 mass of $\Ups$, according to eq.(9.5) of \cite{pdg04}".\par
  Line 1 of Part II, ``Exact APT", presents results of
 $\Ups$1S decay calculated by exact numeric tables for
 $\agoth_3\,$ and by \cite{MagrPriv} for $\agoth_4\,$ APT
 function.
 Line 2, [Mod], presents values obtained from $\Ups(1S)\,$
 decay data by means of the Model eqs.(12),(13). Here, model
 errors combine Model errors of both the terms in the r.h.s.
 of eq.(15). Line 3, ``Crude APT" gives an earlier
 result\cite{Sh01epjc}  with crude APT estimate used to correct
 the  Bethke-2000 value $\as(M_\Ups)=0.170\,$ extracted there
 from all the $\Ups$ decays data. \par

\section*{Acknowledgements}
  An essential part of this material is based on common work
 with Andrej Zayakin. The author is grateful also to A.Bakulev,
 A.Kataev, S.Mikhailov, A.Pivovarov for important advices, as
 well as to Roman Pasechnik for help in numerical calculation.
 The work has been supported in part by RFBR grant No. 05-01-00992
 and by Scient.School grant 2339.2003.2.


\small
\begin{thebibliography}{99}
  \bibitem{Zay05}  D.~V.~Shirkov and A.V. Zayakin;
 hep-ph/0512325.
\bibitem{Bethke00} S.~Bethke, hep-ex/0004021, J. Phys.
  G, {\bf 26} R27 (2000); S. Bethke, hep-ex/0407021,
 Nucl.Phys.Proc.Suppl. {\bf 135}, 345 (2004).
\bibitem{pdg04} S.~Eidelman {\it et al.}, Phys.\ Lett.
       B {\bf 592}, 1 (2004).
\bibitem{badri00} B.~A.~Magradze,``QCD coupling up to third order in standard
 and analytic perturbation theories", RMI-000-15 (2000);
 JINR-E2-2000-222 (2000); hep-ph/0010070.
\bibitem{Sh01epjc} D.~V.~Shirkov, hep-ph/0107282;
  Eur.\ Phys.\ J.\ C {\bf 22}, 331 (2001).
\bibitem{Sh05bari}  D.~V.~Shirkov, in {\it QCD@Work 2005
 (Proceed. Int'nl Workshop on QCD, Conversano, Bari,
 July '05}, edited by P.Colangelo et al., AIP Conf.
 Proceed. {\bf 806}, pp. 97-103; hep-ph/0506050;
\bibitem{Sh01tmp} D.~V.~Shirkov, hep-ph/0012283;
  Theor.\ Math.\ Phys.\  {\bf 127}, 409 (2001).
 \bibitem{KourMagr01}
  D.~S.~Kurashev and B.~A.~Magradze, hep-ph/0104142;
  Theor.\ Math.\ Phys.\  {\bf 135}, 531 (2003)
\bibitem{magr03} B. A. Magradze,``Practical techniques
 of analytic perturbation theory of QCD",
 Report RMI-2003-05; hep-ph/0305205.
  \bibitem{ShSolPL97}
  D.~V. Shirkov and I.~L. Solovtsov, hep-ph/9711251;
  Phys. Lett. B {\bf 442}, 344 (1998).
\bibitem{KrasPiv82}  N.~V. Krasnikov, A.~A.~Pivovarov,
   Phys. Lett. B    {\bf 116}, 168 (1982).
\bibitem{Olga03} O.~P.~Solovtsova,
  Theor.\ Math.\ Phys.\  {\bf 134}, 365 (2003).
\bibitem{BP-KSS04}
 A.~P.~Bakulev {\it et al.},  hep-ph/0405062;
  Phys. Rev. D {\bf 70}, 033014 (2004).
\bibitem{BMS05}
 A.~P.~Bakulev, S.~V.~Mikhailov, and N.~G.~Stefanis, hep-ph/0506311;
 Phys. Rev. D {\bf  72},  074014 (2005);
 A.~P. Bakulev, A.~I. Karanikas, and
   N.~G. Stefanis, hep-ph/0504275;
  Phys. Rev. D {\bf 72}, 074015 (2005).
 \bibitem{Milan02-4} M. Baldicci and G.~M. Prosperi,
 Phys. Rev. D {\bf 66}, 074088 (2002); hep-ph/0412359,
 in {\it  AIP Conf. Proc.} {\bf 756}, pp. 152-161, (2005).
\bibitem{SS-99}  I.~L.~Solovtsov and D.~V.~Shirkov,
 hep-ph/9909305; Teor.\ Mat.\ Fiz.\  {\bf 120} (1999) 482;
  [Theor.\ Math.\ Phys.\ {\bf 120}, 1220 (1999)].
\bibitem{Bodwin95} G.~T.~Bodwin, E.~Braaten, and
 G.~P.~Lepage, hep-ph/9407339;
  Phys.\ Rev.\ D {\bf 51}, 1125 (1995)
\bibitem{Mackenz81}  P.~B. Mackenzie and G.~P. Lepage,
 Phys.Rev.Lett., {\bf 47}, 1244 (1981).
\bibitem{Adams:2004xa} G.~S.~Adams {\it et al.}
 [CLEO Collaboration],hep-ex/0409027;
 Phys. Rev. Lett.  {\bf 94}, 012001 (2005).
\bibitem{MagrPriv}
B. Magradze, {\it private communication}.
\end{thebibliography}
\end{document}